\documentclass[usenatbib]{mnras}

\usepackage{newtxtext,newtxmath}

\usepackage[T1]{fontenc}

\DeclareRobustCommand{\VAN}[3]{#2}
\let\VANthebibliography\thebibliography
\def\thebibliography{\DeclareRobustCommand{\VAN}[3]{##3}\VANthebibliography}

\usepackage{graphicx}	
\usepackage{amsmath}	

\title[The curse of clouds]{The curse of clouds: overcoming challenges of exoplanet spectroscopy}

\author[J. K. Barstow]{
Joanna K. Barstow,$^{1}$\thanks{E-mail: jo.barstow@open.ac.uk}
\\
$^{1}$School of Physical Sciences, The Open University, Walton Hall, Milton Keynes, MK7 6AA\\
}

\date{Accepted XXX. Received YYY; in original form ZZZ}

\pubyear{2021}

\begin{document}
\label{firstpage}
\pagerange{\pageref{firstpage}--\pageref{lastpage}}
\maketitle

\begin{abstract}
In recent years, a vast increase in spectroscopic observations of transiting exoplanets has for the first time allowed us to search for broad trends in their atmospheric properties. Analysis of these observations has revealed that, even for the highly irradiated hot Jupiters, aerosol is a common presence and must be accounted for in modelling efforts. An additional challenge for hot Jupiters is the large variation in temperature across the planet, which is likely to result in partial or patchy cloud cover. As our observational capability is due to increase further with the launch of the \textit{James Webb Space Telescope}, anticipated in autumn 2021, community efforts are underway to prepare modelling and analysis tools capable of recovering information about variable and patchy cloud coverage on hot exoplanets. 
\end{abstract}




\section{Introduction}
Since the first observation of a transiting exoplanet atmosphere in 2002 \citep{charbonneau02}, we have uncovered details of the chemistry and structure for several tens of objects. The majority of these are hot Jupiters - gas giant planets in close orbits around their parent stars, that experience extreme levels of irradiation. 

\begin{figure*}
	\includegraphics[width=\textwidth]{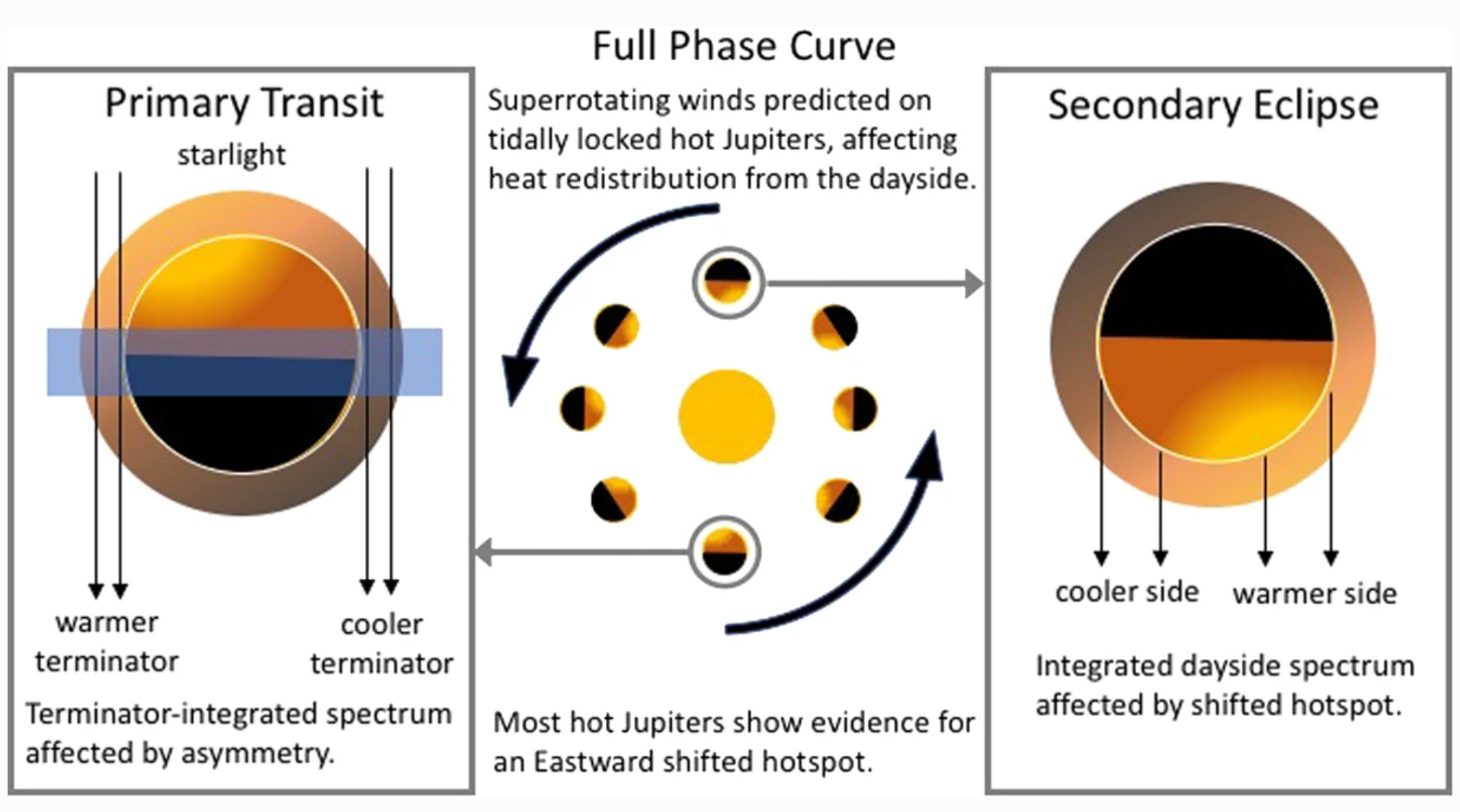}
    \caption{This schematic shows a typical hot Jupiter with an eastward-offset hot spot and superrotating winds, as observed during transit (left), eclipse (right) and over a full phase curve (centre). The phase curve observation provides direct information about the spatial variability of the atmospheric temperature, whereas the transit and eclipse observations average over the terminator and dayside respectively. This figure is reproduced from \protect\cite{barstow20c} with permission from Springer.}
    \label{fig:phase_curve}
\end{figure*}

The most successful method of characterising transiting exoplanet atmospheres has been the transit/eclipse spectroscopy technique (Figure~\ref{fig:phase_curve}). When the planet transits its parent star, a small fraction of the starlight is filtered through the planets atmosphere, emerging with the fingerprints of any absorbing atmospheric gases. When the planet is in turn eclipsed by the star, a measurement of the drop in flux as a function of wavelength reveals the spectrum of light reflected and emitted by the planet itself. 

The planets that have been characterised in most detail by this method are the hot Jupiters. These are gas giant planets, roughly the size of Jupiter, in very close orbits around their parent stars; typically, orbital periods are only a few days. Because of this, the hot Jupiters experience extreme levels of stellar irradiation, and can reach temperatures exceeding 2000 K. 

For the most optimal targets, the changing flux from the planet as a function of phase and wavelength can be obtained, allowing the variation of atmospheric features to be mapped. For hot Jupiters WASP-43b \citep{stevenson14}, WASP-103b \citep{kreidberg18} and WASP-18b \citep{arcangeli19}, spectroscopic phase curves reveal the changing vertical thermal structure as a function of longitude. These, and other hot Jupiters for which photometric phase curves are available, exhibit hot spots with offsets from the substellar point that range from 0 to 40$^{\circ}$ in longitude. 

These temperature variations are likely to impact other atmospheric properties, most notably condensational clouds. Different species expected to occur in exoplanet atmospheres condense out at different temperatures \citep{wakeford17}, and for planets with extreme temperature variation as a function of phase this is predicted to lead to variable cloud coverage around the planet \citep{parmentier16}. Variable cloud coverage, especially around the terminator region that we probe during transit, can present challenges to the interpretation of spectroscopic data \citep{line16}. 

\subsection{Transit spectra of hot exoplanets}
Currently, the majority of transit, eclipse and phase curve spectra are obtained using the \textit{Hubble Space Telescope} (\textit{HST}). Hubble can be used to observe exoplanet atmospheres from the near ultraviolet through to the near infrared, the majority of observations being made with the Space Telescope Imaging Spectrograph (STIS) and Wide Field Camera 3 (WFC3) instruments and spanning wavelengths from 0.3 --- 1.6 $\upmu$m. Observations are also possible from the ground, for example using the FORS2 instrument on the \textit{Very Large Telescope}.

Whilst our current wavelength coverage is limited, it has enabled measurements of water vapour abundance (e.g. \citealt{wakeford18}); reflected starlight from the dayside of a planet \citep{evans13}; temperature structure of the lower atmosphere (e.g. \citealt{stevenson14b}); and absorption due to metallic species such as sodium (e.g. \citealt{vidal-madjar11}, \citealt{nikolov18}). However, the launch of the \textit{James Webb Space Telescope} (\textit{JWST}), due in October 2021, will grant us wavelength coverage further into the infrared, allowing us to access absorption features for multiple molecules that are currently challenging or inaccessible. \textit{JWST} will also have significantly higher signal to noise, due to its 25 m$^2$ primary mirror, which will allow us to view transiting exoplanet atmsospheres in greater detail than is currently possible. 

\subsection{Temperature variation and bias}
A key impact of extremely non-uniform temperatures is variation in the physical thickness of the atmosphere. In hydrostatic equilibrium, the rate at which atmospheric pressure $p$ decreases as a function of altitude $z$ is determined by the atmospheric scale height, $H$:

\begin{equation}
    p(z)=p(0)e^{-z/H}
\end{equation}

A smaller scale height leads to pressure dropping off more rapidly with altitude, and results in a thinner atmosphere. The scale height $H$ depends on the mean molecular weight of the atmosphere $\mu$, the gravitational acceleration $g$ and the atmospheric temperature $T$:

\begin{equation}
    H = \frac{kT}{{\mu}g}
\end{equation}

where $k$ is the Boltzmann constant. Thus, a higher temperature results in a larger scale height and a more extended atmosphere. 

\cite{caldas19}, \cite{macdonald20} and \cite{lacy20} all consider the impact of this effect on transmission spectra. Transit spectra are primarily sensitive to the region of the atmosphere called the terminator - the division between day and night. \cite{caldas19} find, considering only a day-night temperature gradient with an associated variation in atmospheric thickness, that the temperature recovered from a retrieval analysis of the spectrum is biased towards the hotter dayside temperature from the true terminator temperature. The water vapour abundance is also overestimated. Conversely, \cite{macdonald20} find that not accounting for temperature differences between the east and west terminator regions results in the inference of cooler temperatures than expected; like \cite{caldas19}, they also find that water vapour abundance is typically overestimated.

\subsection{Aerosols in transmission spectra}
\label{sec:cloud_trans} 

Aerosols can have a particularly dramatic effect on exoplanet transmission spectra. In transmission, starlight travels a path through the atmosphere tangential to the surface of the planet, meaning that if any cloud or haze is present the atmosphere rapidly becomes opaque below the top of the cloud. 

\begin{figure*}
	\includegraphics[width=\textwidth]{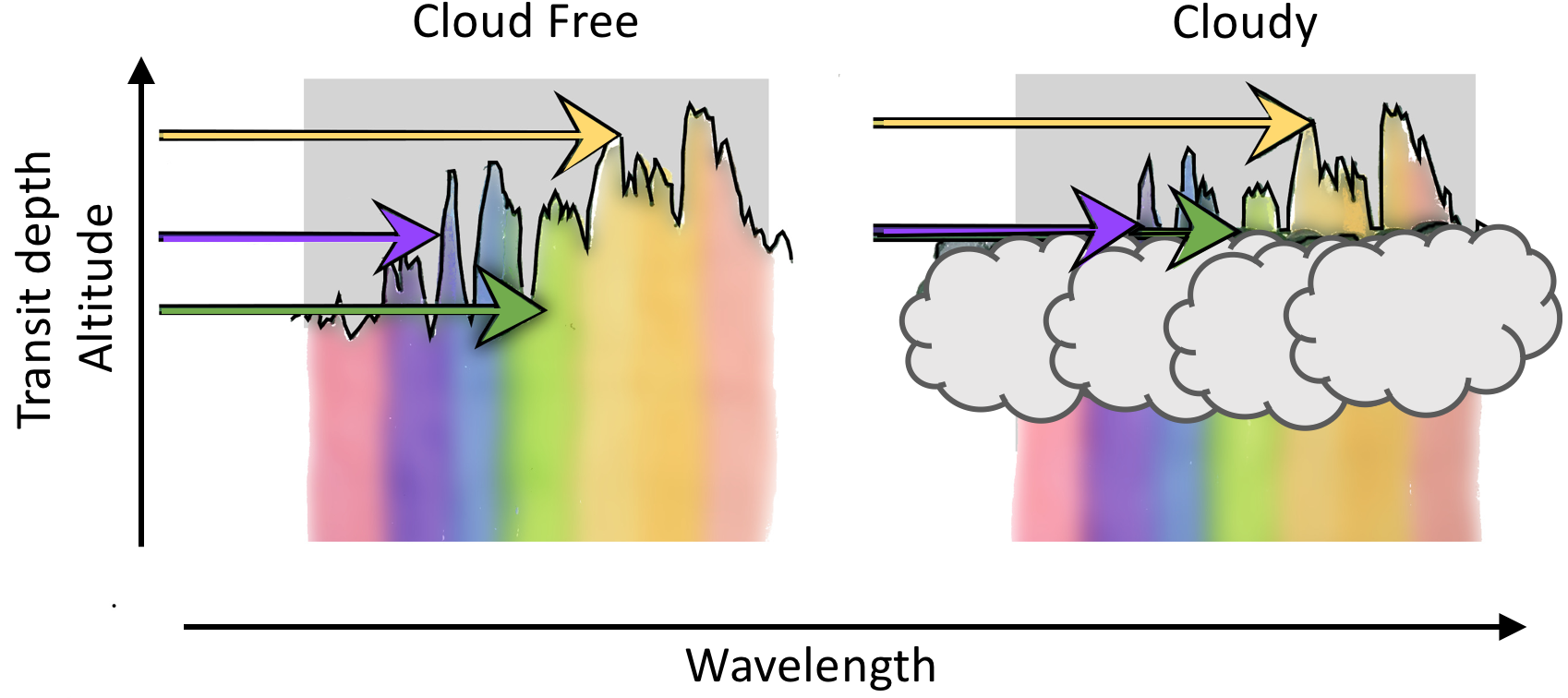}
    \caption{This figure illustrates the effect of clouds on the transmission spectrum of a planet. The black line in each panel shows how the transit depth varies as a function of wavelength; larger transit depths correspond to starlight being absorbed at higher altitudes within the atmosphere. Different wavelengths are indicated by multicoloured shading, and photons at particular wavelengths are shown as coloured arrows. The penetration altitude of each photon within the atmosphere is shown by the vertical position of the arrow. The presence of clouds in the right hand panel can be seen to flatten and raise the lower part of the spectrum, and prevents photons from reaching the deeper regions of the atmosphere. }
    \label{fig:cloud_transit}
\end{figure*}

This impacts our ability to accurately constrain the abundances of molecules within the planet's atmosphere. A cloud deck that is sufficiently high up in the atmosphere prevents the starlight from passing through the deeper regions of the atmosphere, so we lose information about the more transparent edges of the molecular absorption features  (Figure~\ref{fig:cloud_transit}). This removal of the baseline makes it harder to relate the feature amplitude to an abundance. In some extreme cases, such as that of the mini Neptune GJ 1214b, the cloud deck is so high up in the atmosphere that the molecular features in the spectrum are wiped out completely \citep{kreidberg14}.

This picture is complicated further if the terminator cloud coverage is non-uniform. In this instance, the measured spectrum is an average of the clear and cloudy cases. If fractional cloud coverage is not accounted for within models, this could easily be misinterpreted (e.g. \citealt{line16}), leading to inaccurate measurements of molecular abundances. 

With the launch of \textit{JWST}, our sensitivity to these biases is only going to increase. \cite{lacy20} conduct an investigation into the effects of cloud and haze on biases introduced by inhomogeneous atmospheric temperature. They simulate a range of planets as observed by \textit{JWST}, and they find that the presence of high altitude haze especially exacerbates the effect of temperature gradients. 

\section{Atmospheric modelling and retrieval}
It is therefore necessary to ensure that clouds are adequately represented within the models we use to interpret observational data. The modelling tool of choice for this is often a retrieval model, which consists of a relatively simple, parameterised model atmosphere (usually 1D) coupled to an algorithm that samples the available parameter space at random and computes a likelihood for each combination of parameters. Popular sampling approaches for exoplanet applications include Markov-Chain Monte Carlo and Nested Sampling. 

Several exoplanet retrieval tools exist (e.g. NEMESIS \citep{irwin08}; POSEIDON \citep{macdonald17}; TauREx \citep{waldmann16}; ARCiS \citep{min20}; HELIOS-R \citep{lavie17}; ATMO \citep{wakeford17b}; CHIMERA \citep{line14}; SCARLET \citep{benneke15}; petitRADTRANS \citep{molliere20}; AURA \citep{pinhas18}; HyDRA \citep{gandhi18} and PLATON \cite{zhang19}), with atmospheric models that vary in complexity from those that assume self-consistent equilibrium chemistry, to those with freely-varying molecular abundances decoupled from the atmospheric temperature. The extent to which the model is coupled to physical assumptions represents a trade off between allowing prior knowledge of physics to inform and constrain the solution, against allowing the model to freely fit the data and, potentially, highlight inadequacies in our understanding of the underlying physical processes. 

Another consideration in developing such models is the requirement to keep the number of variables to a minimum, in order to avoid overfitting. This is particularly critical with current observations from the \textit{HST} since spectra can have as few as 10 spectral points. It is a particular challenge when including aerosols, which are complex phenomena; properties such as composition, size distribution, location and number of particles can all affect the spectrum we observe, and many of these properties can only be fully described with multiple variables.

The atmospheric retrieval process begins with the definition of a simple model atmosphere. This will usually involve 1) some specification for temperature as a function of pressure; 2) abundances of a range of molecules, with absorption as a function of wavelength for each; 3) bulk properties of the planet such as mass and radius; and 4) properties of any aerosol present. Not all of these parameters will necessarily be variables - for example, if the mass of the planet is known from radial velocity measurements then the mass will be fixed. Generally, abundances of any major atmospheric constituents with minimal absorption features, such as hydrogen and helium, will also be fixed. 

Once the variables have been determined, the prior range for those variables must be specified. We still have a great deal to learn about the atmospheres of exoplanets, so the priors in most cases are very broad and uniform to allow the retrieval maximum freedom to explore the parameter space. This is in contrast to Solar System planets, for which priors are often informed by previous measurements or missions and can legitimately provide much tighter constraints. 

After the initial setup, the retrieval algorithm will randomly draw a set of values from the prior distribution for the variable model parameters. These will set up the model atmosphere for a radiative transfer simulation, called the forward model, which will calculate a spectrum based on that iteration of the atmospheric state. This spectrum is compared with the observation, and a likelihood value assigned to that set of model parameters based on how closely the two match. The process then repeats, with solutions having higher likelihood retained and the sampled parameter space gradually shrinking around the likelihood maximum. The result is a joint probability distribution for the model variables. 

\subsection{The NEMESIS retrieval suite}
NEMESIS \cite{irwin08} was originally developed to analyse spectra of Saturn from the \textit{Cassini}/CIRS instrument, before being expanded to include other Solar System bodies and, more recently, exoplanets and brown dwarfs. It combines a 1D, free-chemistry atmospheric model with the choice of either an optimal estimation \citep{rodg00} or the PyMultiNest \citep{buchner14,feroz08,feroz09,feroz13} nested sampling package, which is more commonly used for exoplanet applications \citep{krissansen-totton18}. NEMESIS also uses the correlated-k approximation \citep{lacis91} to pretabulate the molecular absorption data, which rapidly reduces the computation time for the forward model calculation. 

Because NEMESIS is free-chemistry, no physical constraints are placed on the abundances of the molecular and atomic species. It is therefore possible to retrieve abundances that differ substantially from the expected chemistry. Such results reveal either missing or incorrect opacity within the NEMESIS model, or inadequacies in our physical understanding of the atmospheric system. Comparing results from free retrievals with expectations from chemical models provides us with useful insights that can further our understanding of atmospheric physics. 

\section{Representing clouds in retrieval models}
The challenge for the inclusion of cloud or haze in retrieval models is to represent it in a way that captures the important observable impacts, whilst minimising the number of parameters used. The approach taken will vary depending on the wavelength range covered by the dataset and the geometry of the observation (transit or eclipse).

\begin{figure}
	\includegraphics[width=\columnwidth]{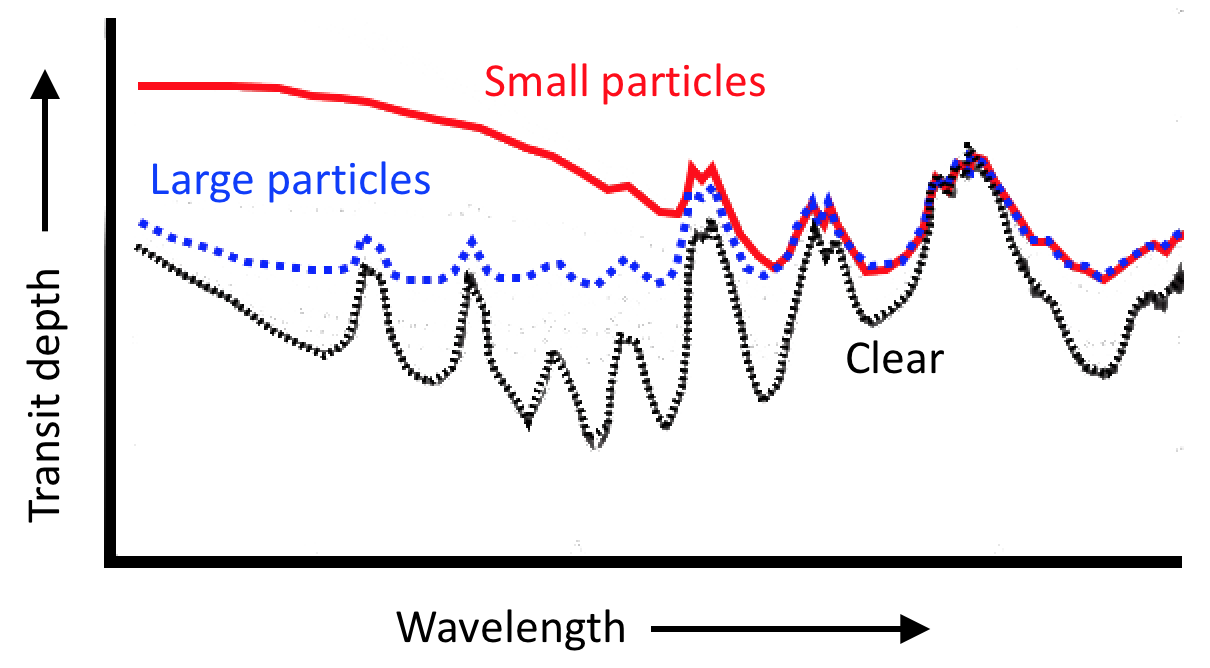}
    \caption{The curves in this figure show transit spectra for a clear atmosphere (bottom), one with large cloud particles (middle) and small cloud particles (top). Whilst both cloud types obscure gas absorption features, the large cloud particles introduce a flat bottom to the spectrum, whereas the small particles create a slope with decreasing cloud opacity as a function of wavelength.}
    \label{fig:cloud_particle}
\end{figure}

In transit, the altitude of the cloud top has a particularly strong effect on the spectrum (see Figure~\ref{fig:cloud_transit}). The higher the cloud top, the more it will obscure the molecular absorption features. Therefore, some parameterisation of the vertical extent and position of the cloud is critical to include. The wavelength dependence of the cloud opacity will also be important; for particles that are small relative to the wavelength of light, the opacity will drop off rapidly as the wavelength increases, whereas for larger particles the opacity will be relatively constant with wavelength (Figure~\ref{fig:cloud_particle}).

The number density of cloud particles also has an impact on the spectrum, as this will influence the point at which the cloud becomes opaque rather than transparent. 

Several studies have been published in recent years which attempt to characterise and compare a population of cloudy hot Jupiters observed with \textit{HST}. Two of these (\citealt{barstow17} and \citealt{pinhas19}) considered data from the STIS and WFC3 instruments, as well as additional data from  the \textit{Spitzer} space telescope; two others (\citealt{fisher18} and \citealt{tsiaras18}) considered only WFC3 spectra. All four papers adopt a minimal, parametric model to represent the effect of cloud and haze on the spectra, with each representation taking a slightly different approach to the problem.

\begin{figure*}
	\includegraphics[width=\textwidth]{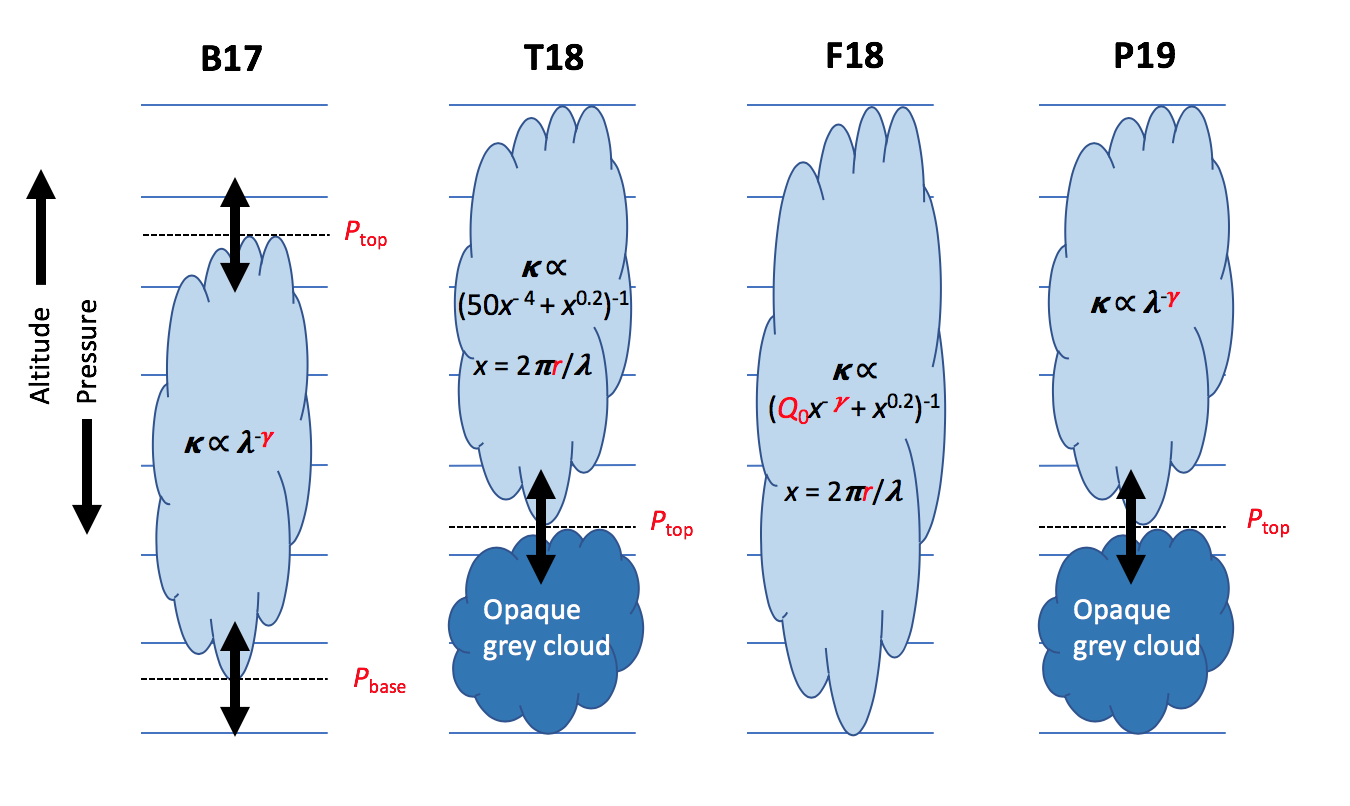}
    \caption{This figure shows the different simple cloud parameterisations that have been used in recent comparative studies of transiting exoplanets. It illustrates how various free parameters, including cloud top and base pressures ($P_{\mathrm{base}}$ and $P_{\mathrm{top}}$), effective particle radius ($r$), scattering index ($\lambda$) and peak extinction factor ($Q_0$) are used to describe the cloud in the four studies \citep{barstow17,tsiaras18,fisher18,pinhas19}.}
    \label{cloud_params}
\end{figure*}

These different approaches are illustrated in Figure~\ref{cloud_params}. Here, the opacity $\kappa$ is a function of wavelength $\lambda$, and may also depend on effective particle radius $r$; a power law index $\gamma$; and $Q_0$, a factor that determines at what wavelength the scattering efficiency peaks. As well as the wavelength dependence of the opacity, a variety of different approaches are used to parameterise the height and vertical extent of the cloud. The model from \cite{barstow17}, referred to as B17 in the diagram, has both variable cloud top and base pressures, whereas the \cite{tsiaras18} and \cite{pinhas19} (T18 and P19) models have two distinct cloud regions, with the lower being a completely opaque grey cloud deck, and the other having wavelength dependent opacity, with the pressure of the transition between them being a free parameter. The \cite{fisher18} model does not have an explicit parameterisation of vertical extent.  

In \cite{barstow20b}, I conduct a comparative study of these different methods for parameterising cloud in retrievals of transmission spectra, by incorporating each of the models into NEMESIS and applying them to the same two datasets; these are the STIS, WFC3 and \textit{Spitzer} observations of hot Jupiters HD 189733b and HD 209458b. These planets are the two best-studied hot Jupiters, and have often been considered the archetypal `hazy' and `cloudy' hot Jupiters respectively. 

These characterisations were borne out by the analysis. Whilst the different cloud models used fit for different variables, a consistent picture emerged for each planet when comparing the results. HD 189733b is best fit by a cloud with small effective particle radius $r$/large power law index $\gamma$, and any grey cloud present is located deep in the atmosphere, whereas the transparent, wavelength dependent portion of the cloud extends to low pressures. The converse is true for HD 209458b, for which the opaque grey cloud dominates for the T18 and P19 models, and larger particle sizes/lower values of the power law index are favoured. This is consistent with expectations that small particles are more easily lofted high into an atmosphere and are therefore more likely to extend to lower pressures, whereas larger particles will sink and are more likely to be situated in the deeper regions. 

Whilst if cloud is not included at all in a retrieval model for a cloudy planet the solution will be biased, reassuringly it seems that provided the effects of the cloud are adequately represented the precise form of the representation doesn't affect the retrieval of properties such as the water vapour abundance. This is true at least for the datasets considered by \cite{barstow20b}, although it may break down in the context of higher precision spectra from \textit{JWST}.  

Parametric cloud models are clearly a powerful tool within exoplanet retrieval. They are however somewhat divorced from physical reality; for example, the power law index retrieved for HD 189733b takes values between 6 and 10. A purely Rayleigh scattering cloud would have an index of 4. Whilst for real clouds 'super Rayleigh' behaviour can occur over a narrow wavelength range (see \citealt{pinhas17}), there is not an obvious mechanism to produce this over the full STIS---Spitzer range. Likewise, whilst the F18 parameterisation (based on the work of \cite{kitzmann18} and previously \cite{lee13}) does encode some indications of composition via the $Q_0$ parameter, for the parameterised retrieval process doesn't make any physical assumptions about the availability of a suitable condensate. Therefore, gaining detailed understanding of the physical properties of cloud requires more than just a parametric model framework. 

\section{Physical cloud models}
Parametric retrieval models are at the simple end of a continuum of cloud models with varying complexity. The degree to which a model contains detailed physics is generally a trade off against computation time and flexibility. It is not feasible, as yet, to couple fully consistent microphysical cloud models to an inversion algorithm, since retrievals using Markov-Chain Monte Carlo or nested sampling techniques generally require several tens of thousands of individual model calculations as a minimum, so a single forward model run must be rapid.

\subsection{The Ackerman and Marley model}
Approaches such as that of \cite{ackerman01} provide a good compromise. This model incorporates a simplified representation of condensation and sedimentation/rainout of cloud particles, which can be described with only a few variables. Models of this type can be used in conjunction with retrieval algorithms as the computation time is still relatively rapid; at the expense of having to make some assumptions, the results from a retrieval analysis can be directly related to physical processes within the atmosphere. 

The Ackerman and Marley (hereafter AM01) model works by assuming that all excess vapour beyond the saturation vapour pressure will condense. In reality, saturation above the saturation vapour pressure is often required for condensation to take place as the curvature of a liquid droplet represents an additional energy barrier to condensation due to increased surface tension for small particles. However, this is to first order a reasonable approximation. 

The vertical extent of the cloud deck and the size of the cloud particles above the base of the cloud are determined by the balance of upward vertical mixing via eddy diffusion, and downward sedimentation of the particles. The motion of the particles is described by the equation

\begin{equation}
    K\frac{\partial q_t}{\partial z}-f_{\mathrm{rain}}w_{\textasteriskcentered}q_c = 0
\end{equation}

where $K$ is the eddy diffusion coefficient, $w_{\textasteriskcentered}$ is the convective velocity scale $f_{\mathrm{rain}}$ is the sedimentation efficiency, $z$ is the altitude, and $q_t$ and $q_c$ are the total and condensate mole fractions of the condensing species. 
The distribution of particle sizes in the model is represented by a log-normal distribution. The geometric mean radius of the distribution scales with $f^{1/\alpha}_{rain}$. The power law index $\alpha$ takes values between 1 and 2 can be defined by fitting the particle fall speeds for different radii around the point where the velocity is equal to the convective velocity scale $w_{\textasteriskcentered}$; in practice, it is often held fixed to an intermediate value such as 1.4 \citep{charnay18}. 

\begin{figure}
	\includegraphics[width=\columnwidth]{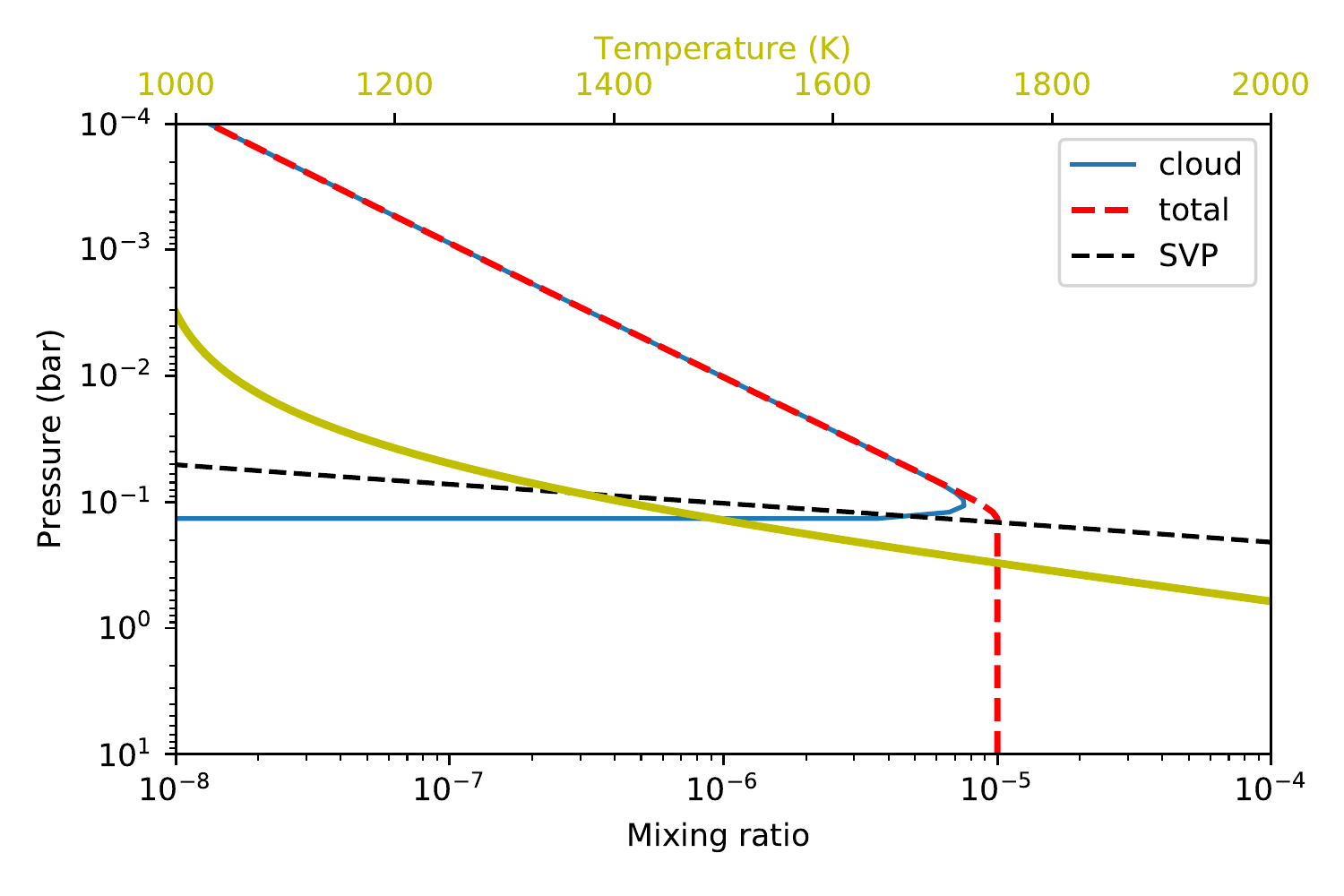}
    \caption{This plot shows the structure of a cloud generated using the AM01 model prescription. The temperature-pressure profile is shown in yellow, and the total mixing ratio of the condensate is shown in red. The black dashed line is the saturation vapour pressure. The cloud starts to form where the condensate mixing ratio crosses the saturation vapour pressure curve. The mixing ratio of the condensed cloud is shown in blue. The mixing ratio of the condensate is constant below the cloud, but decreases with decreasing pressure above the base of the cloud due to sedimentation of the cloud particles. This figure is reused from \protect\cite{yang20} with the permission of author.}
    \label{am01_cloud}
\end{figure}

A cloud deck that is self-consistent with the temperature profile and condensate abundance can thus be calculated using this method with as few as two extra free parameters, $K$ and $f_{rain}$. A typical cloud deck generated using the AM01 model is shown in Figure~\ref{am01_cloud}. This does however require some further underlying assumptions. For example, the saturation vapour pressures of likely condensates must be calculated and included, and likely condensates must also be included in the model atmosphere in vapour form. In some cases, the condensate vapour might have spectral features within the wavelength range studied, but this will not be universally true; therefore independent constraint on the amount of vapour available for condensation may not be possible. Nonetheless, the model represents a good compromise between minimal free parameters, relatively rapid computation time, and a basis in atmospheric physics. 

\cite{mai19} explore the application of different parameterised cloud models to simulated \textit{JWST} datasets. They generate the datasets using the AM01 cloud model, and explore the extent to which simpler, less physical parameterisations can recover the key characteristics of the cloud. They find that the atmospheric gas phase composition is relatively robust to different cloud models, as in \cite{barstow20b}, but unsurprisingly that the recovery of cloud properties is very dependent on the choice of cloud model. They recommend that the AM01 cloud model is used in all cases where the goal is to obtain information about the cloud itself. 

\subsection{Microphysical and kinetic simulations}
It is also possible to simulate the microphysical processes of cloud formation, and generate synthetic spectra based on these models. The computation time prohibits these models from being coupled to a retrieval algorithm; however, as a purely forward modelling tool these simulations can be very informative. 
\cite{gao20} apply the Community Aerosol Radiation Model for Atmospheres (CARMA) to hot Jupiter atmospheres, to investigate haze and cloud formation over a range of temperatures. CARMA explicitly models the nucleation of cloud particles, both homogeneous (self-nucleation) and heterogeneous (nucleating onto a particle of another substance); condensational growth and evaporation; and coagulation. It also deals with the vertical transport of the particles that are formed. This process allows the cloud formation to be modelled from first principles, and does not require assumptions to be made about parameters that do not have a direct physical meaning. 

Whilst microphysics models are too complex and time consuming to be used in a retrieval context, their results can still be compared with available datasets. \cite{gao20} examine the formation of cloud and haze on planets at different temperatures, and compare the results to the observed amplitude of the H$_2$O absorption feature in the \textit{Hubble}/WFC3 bandpass. The amplitude of this feature is affected by the presence of cloud and haze, as explained in Section~\ref{sec:cloud_trans}. For very cloudy atmospheres, the feature can be almost non-existent, so the planets with the lowest amplitude water features are likely to be cloudy. By contrast, planets with semi-transparent high altitude haze might still have substantial water vapour absorption features, depending on the size of the haze particles; if the particles are small, the extinction efficiency may be relatively small in the WFC3 bandpass. 

\cite{gao20} find that the spectra of hot Jupiters with temperatures ranging from less than 1000 K to over 2000 K are mostly dominated by silicate clouds, with hydrocarbon hazes playing a part for only the coolest planets. They find that cloud coverage varies considerably as a function of temperature, with the smallest H$_2$O absorption features in the modelled spectra occurring at temperatures around 1700 K. This matches the feature amplitudes from observational data remarkably well. The largest cloud particles are formed at temperatures of around 1500 K. 

\begin{figure*}
	\includegraphics[width=\textwidth]{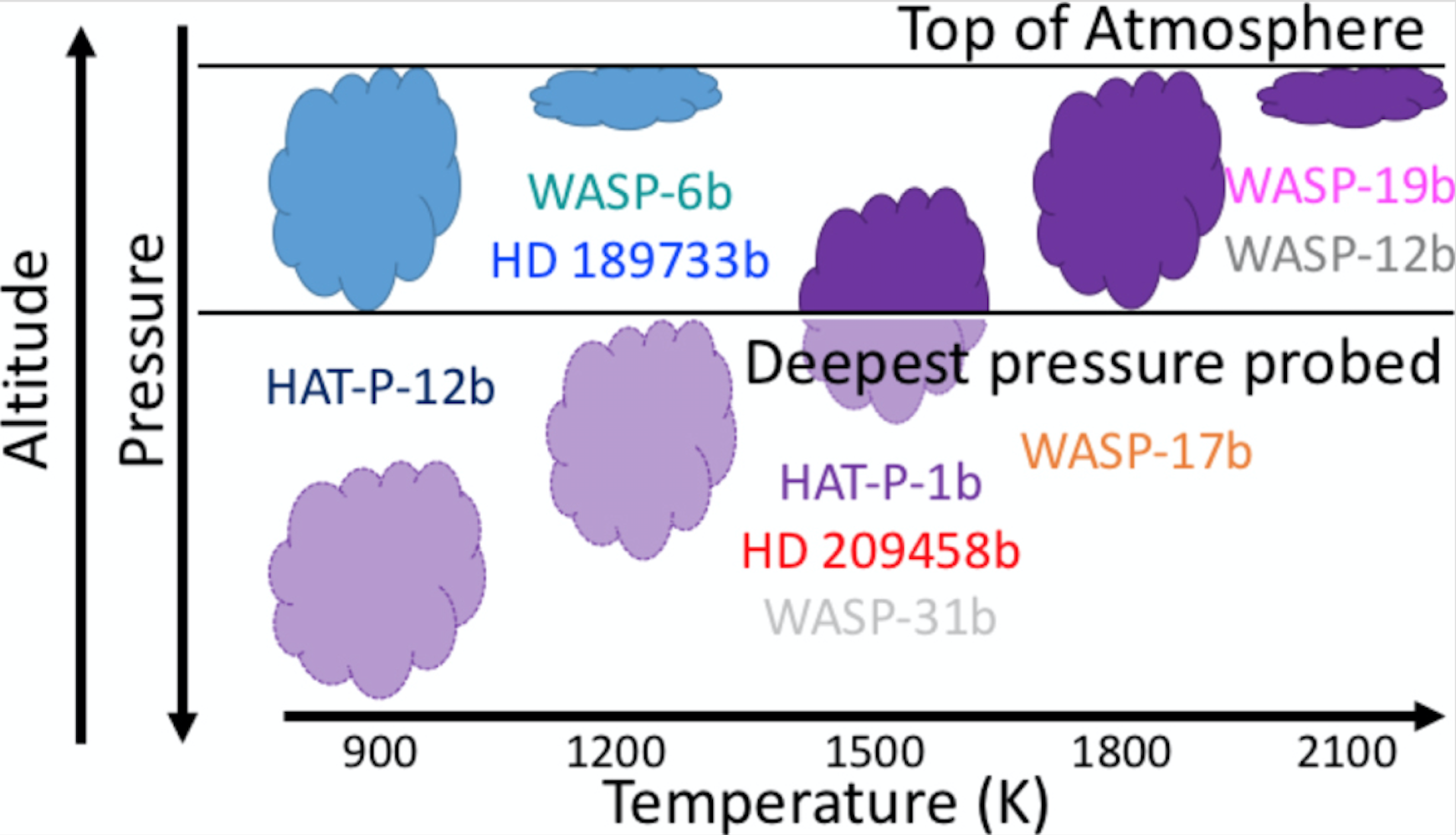}
    \caption{This figure from \protect\cite{barstow17} presents a hypothesis that can explain the retrieved cloud top pressures, and the preference for Rayleigh vs grey wavelength dependence, from the study. The two different colours represent two different condensibles that form cloud at different temperatures. As planets become hotter, any particular cloud will form higher up in the atmosphere assuming the temperature decreases with altitude. At high enough temperatures, some clouds will be unable to condense, and their place will be taken by cloud species with a higher vaporisation temperature.}
    \label{b17cloud}
\end{figure*}

The results from the \cite{gao20} model echo the findings of \cite{barstow17}. In Figure~\ref{b17cloud}, taken from the B17 paper, we see that the retrieved characteristics of cloud can be explained by two different types of cloud, which condense at different temperatures. As planet temperature increases, a given cloud will be forced to form at higher regions in the atmosphere where it is cooler, until it is unable to condense at all. In \cite{gao20}, the species responsible for this sort of pattern are identified as hydrocarbon haze and silicates. 

Larger cloud particles are expected to reside deeper in the atmosphere, and indeed in B17 we found that the deepest visible clouds were found on planets with temperatures around 1500 K. This corresponds to the temperature at which the cloud particle size is maximised in the \cite{gao20} study. This agreement between observation-driven and detailed modelling studies is very encouraging; it indicates that our understanding of the physics of cloud formation in hot atmospheres is reasonably correct. It also highlights the importance of applying models of various levels of complexity to the problem.

Other groups have also made use of this type of modelling. A considerable body of work on dust grain formation in brown dwarf and stellar atmospheres is encapsulated in the development of the DRIFT code \citep{woitke03, woitke04, helling06, helling08b}. This code considers the formation of condensation nuclei that might facilitate the formation of clouds in hot atmospheres, as well as the formation of dust and clouds themselves. Elements of the DRIFT code have also been incorporated into global climate models, which allows the 3D effects of cloud to be investigated.

\subsection{Cloud in Global Climate Models}
In recent years, clouds have begun to be incorporated into Global Climate Models (GCMS). In some cases, this has been simply as a passive tracer, whilst in others the radiative feedback effects of the cloud have also been included. 

\cite{parmentier16} use the results from an existing 3D GCM to predict where various clouds will form in an exoplanet atmosphere. Clouds form where the partial pressure of the condensate exceeds the saturation vapour pressure for that species. The cloud particle size, and the location of the cloud top, are both free parameters within the model; sedimentation and growth processes are not considered. 

This equilibrium cloud condensation model indicates that different types of cloud that form at different temperatures would affect the shape of an optical phase curve of the planet, with the peak occurring at different phases depending on what type of cloud has formed. This is due to the large temperature variation as a function of phase, which means that cloud forms in some locations but not in others. \cite{parmentier16} suggest that the location of the optical phase curve maximum can be used as a diagnostic for particular cloud species. 

\cite{lee16} incorporate the DRIFT kinetic cloud model into a 3D radiative hydrodynamic simulation of the atmosphere of hot Jupiter HD 189733b. In this case, the cloud is fully coupled to the dynamics and radiative transfer; the cloud formation is governed by the local thermal conditions, and in turn the radiative properties of the cloud feed back into the temperature structure. The simulation reveals a cloud concentrated in the nightside equatorial regions of the planet. The deeper regions of the cloud are dominated by magnesium silicate, but at higher altitudes uncovered seed particles of titanium dioxide are present.

\cite{lines18} build on the work of \cite{lee16} and incorporate key elements of the DRIFT code into the Unified Model, which has been expanded in recent years to simulate hot Jupiter spectra. This model incorporates cloud formation and radiative feedback, and the authors also generate simulated transmission spectra for a variety of modelling scenarios that can then be compared with observations. 

Simulated observations can also be used to generate datasets for planets as they would be seen for future instruments. As the community prepares for the launch of \textit{JWST}, and with increasing awareness that the 3D structure of exoplanet atmospheres is likely to significantly influence what we see, the development of tools such as this is especially timely. 

\section{Exoplanet clouds in the \textit{JWST} era}
Given the expectation that \textit{JWST} will be a game changer for the observation of exoplanet atmospheres, a substantial amount of effort has been invested in simulating \textit{JWST} spectra. This enables the community to make informed decisions about which targets to observe, and to test the tools we plan to use for interpretation. 

Some early efforts to simulate \textit{JWST} spectra, by \cite{barstow15} and \cite{greene16}, used simple 1D parameterised atmospheric retrieval models to test what sort of information could be recovered from observations. These did not include complex or sophisticated cloud models; \cite{barstow15} considered mostly cloud-free atmospheres, with the exception of a single simulation for GJ 1214b, and \cite{greene16} included examples with only a simple opaque grey cloud deck, which was assumed to cover the planet. 

\cite{wakeford15} produce synthetic \textit{JWST} spectra for a range of species. Whilst they do not perform a retrieval test to examine the recoverability of information, the simulations reveal that absorption features for multiple cloud species occur towards the longer end of the \textit{JWST} wavelength range. Therefore, with \textit{JWST} direct detection of particular cloud constituents would be possible. 

Due to the required speed of computation for retrieval models that rely on either MCMC or Nested Sampling for iterating towards a preferred solution, multiple scattering in clouds has generally been ignored for exoplanets. Whilst several examples exist of multiple scattering clouds being modelled in the Solar System, these simulations are usually coupled to an algorithm such as Optimal Estimation \citep{rodg00}, which requires only a few tens of forward modelling runs to arrive at a solution. \cite{barstow14} did model multiple scattering in order to characterise the reflected starlight from hot Jupiter HD 189733b, but only in the context of forward modelling rather than retrieval. 

Indeed, the majority of exoplanet retrieval models adopt an extinction-only approximation. This results in all photons that interact with a cloud particle being either absorbed or scattered out of the beam. In reality, depending on the single scattering albedo (the fraction of light scattered vs absorbed) and the scattering phase function (which direction photons are preferentially scattered in), some percentage of photons that interact may be forward-scattered and remain within the beam, and some that are scattered out may be scattered back into the beam due to interaction with another particle. Full treatment of scattering in models can, therefore, result in very different results to the extinction-only approximation, depending on the characteristics of the cloud. 

\cite{taylor20} investigate the impact of including scattering on eclipse spectra of hot Jupiters as seen with \textit{JWST}. Whilst they do not perform a full multiple scattering calculation, they do consider scattering as distinct from extinction. The major impact of this is that the atmosphere no longer emits as a blackbody. This can lead, for example, to emission features appearing in a spectrum even in the case where the atmosphere is isothermal. 

\cite{taylor20} use their scattering model to investigate the biases that would result from fitting a spectrum of a planet with scattering cloud with a model assuming pure absorption. They find that atmospheric temperatures would be consistently underestimated in this case. They also develop a simple parameterised model that is able to account for different degrees of scattering, by fitting for the single scattering albedo. The single scattering albedo is parameterised as a step function, with a shift between two variable values at a variable wavelength. This is quite representative of the wavelength-dependent single scattering albedo for a typical cloud. 

As well as the need to consider scattering, spatial variation of cloud coverage and temperature is also a key concern for interpreting \textit{JWST} data. \cite{line16} consider the effect of any cloud present covering only a fraction of the terminator region. Subsequent authors have built on this to consider the effect of atmospheric spatial variability on observed spectra. Recently, \cite{lacy20} examined the effect of strong temperature gradients in the presence of cloud and haze on exoplanet transmission spectra. They find that cloud and haze exacerbates the impact on observed spectra of strong day-night temperature gradients, making it even more critical to account for this. 

In \citet{lacy20b} they further investigate the possibility of recovering information about cloud from \textit{JWST} spectra, using relatively simple cloud models where the cloud is treated either as 1) a well mixed slab up to a certain top pressure, or 2) an equilibrium cloud that condenses once saturation is reached, with a number of particles that drops off towards lower pressures according to a power law with a variable index. Regardless of the model adopted, \cite{lacy20b} find that in many cases the composition of the aerosol can be distinguished from \textit{JWST} spectra, although this is more challenging in the case of different hydrocarbon hazes as their spectra closely resemble each other. 

\section{Conclusions}
The launch of \textit{JWST} is expected to provide considerable insights into cloudy exoplanets. For example, we hope for the first time to obtain observational evidence for the composition of the clouds present in hot Jupiter atmospheres, which the \textit{JWST} wavelength coverage will allow. Characterising cloud more fully across a range of planets with different atmospheric conditions will provide additional information about dynamics and thermal structure, since cloud formation is intrinsically tied to atmospheric circulation and convection. 

A range of modelling approaches will be required to fully exploit these datasets. Interpretation of observations will require retrievals and parameterised models, which will necessarily be more sophisticated than those that are currently available. The recent work highlighted here has provided several possible avenues for this development, with the following factors being key considerations:
\begin{itemize}
\item Spatial variation in cloud coverage
\item Spatial variation in temperature
\item Inclusion of scattering in eclipse spectra
\end{itemize}

3D climate simulations that contain physically motivated cloud models will be an important resource for benchmarking new models and methods (e.g. \citealt{barstow19}). Testing the ability of a retrieval to correctly recover a known solution is an important validation step, and physical simulations inform us about the phenomena we should be able to recover. 

Community level efforts have ensured open access to Cycle 1 data from \textit{JWST} via the Early Release Science programme, and similar strategies are now underway to discuss and implement necessary developments in modelling tools - for example, the January 2021 Specialist Discussion Meeting on Exoplanet Modelling in the \textit{JWST} Era. The next few years will no doubt be an exciting time, with a fair share of new (and perhaps controversial!) discoveries; my hope is that the community will build on the current collaborative spirit as we further our understanding of these fascinating worlds.

\section*{Acknowledgements}
I thank Jingxuan Yang, who recently worked with me as a Masters student, for his excellent work implementing the AM01 model in NEMESIS, and for giving me permission to use an explanatory figure from his thesis. For the majority of my own work that I discuss in the article I was supported by the Royal Astronomical Society Research Fellowship. I am currently an Ernest Rutherford Fellow supported by the Science and Technology Facilities Council.

\bibliographystyle{mnras}
\bibliography{bibliography} 




\bsp	
\label{lastpage}
\end{document}